\documentclass[conference]{IEEEtran}
\IEEEoverridecommandlockouts

\usepackage[T1]{fontenc}
\usepackage[utf8]{inputenc}
\usepackage{cite}
\usepackage{amsmath,amssymb,amsfonts}
\usepackage{graphicx}
\usepackage{textcomp}
\usepackage{xcolor}
\usepackage{booktabs}
\usepackage{array,tabularx}
\usepackage{url}
\usepackage{microtype}
\usepackage[varqu,narrow]{zi4}

\newcommand{\feat}[1]{\texttt{\detokenize{#1}}}

\newcolumntype{L}[1]{>{\raggedright\arraybackslash}p{#1}}
\newcolumntype{Y}{>{\raggedright\arraybackslash}X}

\begin{document}

\title{%
{\normalfont\fontsize{7.5}{8.5}\selectfont
\mbox{This work has been submitted to the IEEE for possible publication. Copyright may be transferred without notice, after which this version may no longer be accessible.}
\par\vspace{2mm}}
From Ethernet to Wi-Fi: Flow Feature Selection
and Construction for Traffic Classification
}

\author{
\IEEEauthorblockN{Francisco Rau\IEEEauthorrefmark{1}\IEEEauthorrefmark{2}, Petia Georgieva\IEEEauthorrefmark{3}, Carlos Herranz\IEEEauthorrefmark{1}, I\~naki Val\IEEEauthorrefmark{1} and Joaquin Perez\IEEEauthorrefmark{2}}
\IEEEauthorblockA{\IEEEauthorrefmark{1}\textit{MaxLinear, Inc.}, Valencia, Spain\\
\IEEEauthorrefmark{2}\textit{Electronic Engineering Department, Universitat de Val\`encia}, Valencia, Spain\\
\IEEEauthorrefmark{3}\textit{Institute of Electronics and Informatics Eng. of Aveiro (IEETA), University of Aveiro}, Portugal\\
Corresponding author: frau@maxlinear.com}
}

\maketitle

\begin{abstract}
Flow-based traffic classifiers trained on Ethernet data may lose accuracy when applied to Wi-Fi traffic. This paper proposes a five-stage procedure for selecting and constructing features that enable Wi-Fi flow classification using models trained only on Ethernet data.
The procedure removes the transport protocol identifier, examines rate dependence, compares feature distributions and the
information each feature provides for the application
classes, reduces redundancy, and constructs
normalized packet-size and temporal features. Ethernet
and Wi-Fi traffic are simulated in ns-3, with Wi-Fi
using Best Effort access category. Labelled Wi-Fi data from six
development scenarios are used for feature design,
while a separate scenario is reserved for final testing.
The evaluation compares feature sets, examines
threshold sensitivity, and analyses classification
errors under different traffic conditions.
\end{abstract}


\section{Introduction}\label{sec:introduction}


The classification of downlink traffic at the Wi-Fi access point (AP) is a critical enabler for delivering high Quality of Service (QoS) in modern wireless networks. Ethernet data is often the preferred choice for building traffic classifiers 
because it provides a controlled and stable environment. 

In our previous work we developed a flow-based traffic classifier to identify applications from packet sizes, packet counts, and inter-arrival times in Transmission Control Protocol (TCP) and User Datagram Protocol (UDP) flows, without inspecting packet payloads \cite{rau}.
However, classifiers trained on Ethernet traffic may lose accuracy
when applied to Wi-Fi. Wireless contention and network
load can change packet timing and feature distributions,
making classification sensitive to differences between
training and deployment conditions~\cite{rosetta}.
Therefore, selecting features that remain informative
across both media is important. Since collecting labelled
traffic under different congested Wi-Fi conditions in
real networks is costly, we use the ns-3 network
simulator to study this problem under controlled conditions.

Previous studies have addressed traffic classification
using different feature selection and representation
methods. Moore and Zuev~\cite{moore} combine a Fast
Correlation-Based Filter with Naive Bayes classification.
FlowPic~\cite{flowpic} represents flows as two-dimensional
histograms of packet size and arrival time and uses a
convolutional neural network. Other studies consider
changes in traffic distributions. Rosetta~\cite{rosetta}
combines TCP-aware augmentation with self-supervised
learning for Transport Layer Security (TLS) traffic
classification and evaluates Ethernet and Wi-Fi.
FG-SAT~\cite{fgsat} uses Jensen--Shannon divergence
to select attributes that preserve class differences
across environments and learns graph-based flow
representations. Yuan et al.~\cite{yuan} use mutual
information and Kullback--Leibler divergence to
evaluate feature information and stability over time
for Tor hidden-service fingerprinting. Their work
focuses on temporal changes rather than differences
between network media.

In this paper, we propose a five-stage procedure to
select and construct features for Wi-Fi traffic
classification using models trained only on Ethernet
data. The procedure examines rate dependence,
distribution differences, class information, and
feature redundancy before constructing normalized
packet-size and temporal features. 
The initial 29 dimensional feature space is reduced to seven input features representative across both media. 
Labelled Wi-Fi data from six
development scenarios (S1--S6) are used for feature
design, based on a classifier trained only with Ethernet
data. A separate Wi-Fi scenario (S7), containing all
traffic classes is reserved for final evaluation.

\section{Methodology}\label{sec:methodology}

\subsection{Data Processing}

Flows are identified by their Internet Protocol (IP) addresses, transport ports, and protocol. For TCP, flow statistics include downlink packets and their corresponding reverse-direction TCP acknowledgments. For UDP, only downlink packets are used. Features are computed from packet metadata in five-second windows advanced every second. Packets without IP payload are discarded, while pure TCP acknowledgments are retained because they contain TCP headers even without application data.\par
The AP-side classifier targets only downlink application flows forwarded by the AP of the home basic service set (BSS) 1. Uplink application traffic is excluded. Observed duration is the interval between the first and last packets in a window. The eight classes, common to Ethernet and Wi-Fi, include Voice over Internet Protocol (VoIP) and video streaming using Dynamic Adaptive Streaming over HTTP (DASH). Table~\ref{tab:apps} lists the classes, Ethernet training rates, and dataset sizes.

\begin{table}[!htbp]
\caption{Common application classes and dataset windows.}
\label{tab:apps}
\centering\footnotesize
\setlength{\tabcolsep}{2.2pt}
\begin{tabularx}{\columnwidth}{@{}l Y c r r r r@{}}
\toprule
 & & \textbf{Rate for training} & \multicolumn{2}{c}{\textbf{Ethernet}} & \multicolumn{2}{c}{\textbf{Wi-Fi}}\\
\cmidrule(lr){4-5}\cmidrule(l){6-7}
\textbf{ID} & \textbf{Application (classes)} & \textbf{(Mb/s)} & \textbf{Train} & \textbf{Test} & \textbf{S1--S6} & \textbf{S7}\\
\midrule
VP & VoIP & 0.10 & 216 & 69 & 1,620 & 270\\
VC & Video Conference & 3 & 216 & 69 & 2,160 & 270\\
GM & Gaming (Quake~3) & 0.11 & 216 & 69 & 1,890 & 270\\
VS & Video Streaming (DASH) & 20 & 216 & 69 & 1,876 & 540\\
BT & Bulk TCP & 10 & 216 & 69 & 742 & 270\\
IT & IPERF TCP & 5--200\textsuperscript{*} & 216 & 69 & 270 & 270\\
IU & IPERF UDP & 10 & 216 & 69 & 1,080 & 270\\
BU & Bursty UDP & 10 & 216 & 69 & 1,080 & 270\\
\midrule
 & Total & & 1,728 & 552 & 10,718 & 2,430\\
\bottomrule
\end{tabularx}
\par\vspace{2pt}
\parbox[t]{\columnwidth}{\footnotesize IDs are reused in Table~\ref{tab:scenarios}; the rates shown are for Ethernet training. \textsuperscript{*}5, 25, 100, and 200~Mb/s, 54 windows each. Wi-Fi counts include five seeds.}
\end{table}

\subsection{Feature Selection and Construction}\label{sec:feature-procedure}

The first four stages apply fixed rules to successively reduce the original 29 features obtained in~\cite{rau} to 28, 16, 12, and 9 features. The fifth stage constructs normalized size and temporal features, ending up with 7 final features.
\subsubsection{Transport Exclusion}\label{sec:transport-exclusion}
Removing the transport identifier prevents direct use of TCP or UDP identity, leaving 28 candidates.
\subsubsection{Rate Dependence}\label{sec:rate-dependence}
For each feature $X$, the absolute Pearson correlation is computed between $X$ and the traffic rate within each of the four variable-rate applications: Bursty UDP, IPERF UDP, Video Conference, and IPERF TCP (Section~\ref{sec:ethdata}). The rate-dependence score is the largest of the four absolute correlations, $s(X)=\max_a|r_a(X)|$. A score near one indicates strong linear dependence on rate in at least one application, rather than discrimination between applications. Undefined correlations from constant features are ignored, and features without a valid score are retained. The other four classes have no controlled rate variation in this analysis.

Twelve features are removed with $s(X)\geq0.97$: forward, reverse, and total packet and byte counts; TCP flag count; rate descriptors; and related traffic-intensity and mean-volume descriptors. Using measured throughput or configured rate gives the same twelve exclusions, leaving 16 features.

\subsubsection{Distribution Shift and Class Information}\label{sec:distribution-shift}
\begin{figure}[t]
\centering\includegraphics[width=\columnwidth]{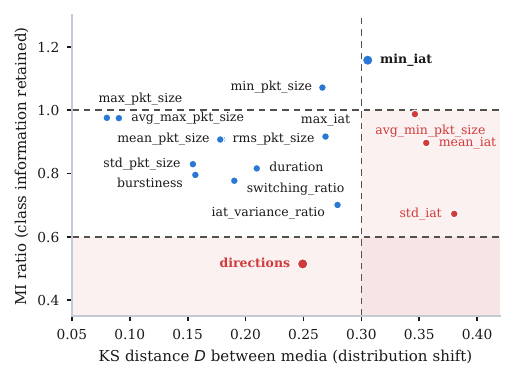}
\caption{Distribution shift and class information for 16 features. Blue: retained; red: removed. Dashed lines show the KS threshold (0.30) and MI-ratio thresholds (0.60 and 1.00). Shaded areas show removal regions.}
\label{fig:transfer}
\end{figure}

For feature $X$ and application label $Y$, the two-sample Kolmogorov--Smirnov (KS) distance~\cite{smirnov1948} is
\begin{equation}
D=\sup_x\left|\widehat{F}_{\mathrm{eth}}(x)-\widehat{F}_{\mathrm{wifi}}(x)\right|,
\label{eq:ks-distance}
\end{equation}
where $\widehat{F}_{\mathrm{eth}}$ and $\widehat{F}_{\mathrm{wifi}}$ are empirical cumulative distribution functions. The distance ranges from 0 to 1 and does not use labels. Mutual information (MI) measures how much information $X$ contains about $Y$~\cite{polyanskiy2025}. We estimate MI separately for each medium with a nearest-neighbor estimator~\cite{kraskov2004,ross2014}. Parameter $k=3$ specifies the three nearest samples used to estimate local probability relationships:
\begin{equation}
\text{MI ratio}=
\frac{\widehat{MI}_{\mathrm{wifi}}(X;Y)}
{\widehat{MI}_{\mathrm{eth}}(X;Y)},
\qquad
\widehat{MI}_{\mathrm{eth}}(X;Y)>0.
\label{eq:mi-ratio}
\end{equation}
A ratio below 1 indicates lower estimated MI on Wi-Fi than on Ethernet, while a ratio above 1 indicates higher MI. Both metrics use 1,728 Ethernet training windows, balanced across eight classes, and 10,718 Wi-Fi windows from S1–S6. These Wi-Fi scenarios deliberately vary application mixes, offered loads, and co-channel contention to represent different congestion conditions. This diversity aims to identify features that remain informative across a broader range of Wi-Fi operating conditions. Thus, KS distance and MI ratio reflect differences in both traffic composition and network conditions, rather than the isolated effect of the transmission medium.
We retain features with MI ratio $\geq0.60$ and either $D\leq0.30$ or MI ratio $\geq1.00$ (Fig.~\ref{fig:transfer}). This removes \feat{avg_min_pkt_size}, \feat{mean_iat}, \feat{std_iat}, and \feat{directions}, leaving 12. Here, \feat{avg_min_pkt_size} is an exponentially weighted moving average of minimum size across consecutive windows of a flow; \feat{directions} is the forward-packet fraction.

\subsubsection{Redundancy Reduction}\label{sec:redundancy-reduction}
Candidates are considered in increasing $D$. A feature is removed if its absolute Pearson correlation with a retained feature exceeds 0.98 in Ethernet training. This retains \feat{max_pkt_size} over \feat{avg_max_pkt_size}, \feat{rms_pkt_size} over \feat{mean_pkt_size}, and \feat{max_iat} over \feat{iat_variance_ratio}, leaving nine. The retained features have lower distribution shift. Here, \feat{avg_max_pkt_size} averages maximum packet size as defined for \feat{avg_min_pkt_size}; \feat{iat_variance_ratio} divides inter-arrival-time (IAT) variance by mean IAT.

\subsubsection{Feature Construction}\label{sec:feature-construction}
We retain \feat{rms_pkt_size}, \feat{max_pkt_size}, and \feat{switching_ratio} and construct four ratio features from the remaining statistics (Table~\ref{tab:ratios}).

\begin{table}[!htbp]
\caption{Rate-dependence scores before and after feature construction ($s(X)$ from Section~\ref{sec:rate-dependence}).}
\label{tab:ratios}
\centering\small
\setlength{\tabcolsep}{3pt}
\begin{tabularx}{\columnwidth}{@{}L{66pt} Y r r@{}}
\toprule
\textbf{Original} & \textbf{Constructed} & $s_{\mathrm{orig}}$ & $s_{\mathrm{constr}}$ \\
\midrule
\feat{std_pkt_size} & \feat{min_max_ratio} & 0.673 & 0.027\\[1pt]
\feat{burstiness} & \feat{pkt_size_range_r} & 0.667 & 0.493\\[1pt]
\feat{max_iat} & \feat{activation_fraction} & 0.803 & 0.803\\[1pt]
\feat{min_iat} & \feat{iat_min_ratio} & 0.803 & 0.766\\
\bottomrule
\end{tabularx}
\end{table}

Both size ratios have lower linear correlation with rate than their original statistics. We additionally use Spearman correlation to compare the ranks of each original statistic and its constructed feature, rather than to test dependence on rate; this comparison does not determine feature selection. For the size ratios, these absolute rank correlations are at least 0.978 in both media. Here, \feat{burstiness} is $(\sigma_s-\bar{s})/(\sigma_s+\bar{s})$, with $\sigma_s$ the packet-size standard deviation. The timing ratios express gaps relative to observed duration or mean gap; \feat{activation_fraction} is smaller when the longest gap occupies more of the observed span.

Table~\ref{tab:features} defines the seven features. Computing them requires nine statistics: root-mean-square (RMS), maximum, minimum, and mean packet sizes; switching ratio; maximum, minimum, and mean inter-arrival times (IATs); and observed duration. The two means, previously excluded as classifier inputs, supply denominators. Switching ratio measures the fraction of consecutive packets with different sizes; \feat{activation_fraction} is the fraction of the observed span excluding the longest gap.

\begin{table}[t]
\caption{Final seven dimensional feature space.}
\label{tab:features}
\centering\small
\setlength{\tabcolsep}{2pt}
\begin{tabularx}{\columnwidth}{@{}L{125pt} Y@{}}
\toprule
\textbf{Feature} & \textbf{Formula}\\
\midrule
\feat{rms_pkt_size} & $\sqrt{\overline{s^2}}$\\[1pt]
\feat{max_pkt_size} & $s_{\max}$\\[1pt]
\feat{switching_ratio} & $n_{\Delta s}/(N-1)$\\[1pt]
\feat{min_max_ratio} & $s_{\min}/s_{\max}$\\[1pt]
\feat{pkt_size_range_r} & $(s_{\max}-s_{\min})/\bar{s}$\\[1pt]
\feat{activation_fraction} & $1-\tau_{\max}/T$\\[1pt]
\feat{iat_min_ratio} & $\tau_{\min}/\bar{\tau}$\\
\bottomrule
\end{tabularx}
\par\vspace{3pt}
\parbox[t]{\columnwidth}{\footnotesize $N$: packet count; $s_i$, $t_i$: packet size and timestamp; $\tau_i=t_i-t_{i-1}$; $T=t_N-t_1$; $n_{\Delta s}$: number of size changes between consecutive packets; $\overline{s^2}=N^{-1}\sum_i s_i^2$. Bars indicate means. Zero denominators are replaced by $10^{-9}$.}
\end{table}

\section{Experimental Setup}\label{sec:setup}

\subsection{Ethernet Datasets}\label{sec:ethdata}

The Ethernet data use an uncongested 1~Gb/s ns-3 link. IPERF TCP uses four training rates and is also evaluated without a rate limit. The balanced test set comes from separate captures with different simulation seeds and includes rates absent from training.

Rate dependence is assessed with 348 additional Ethernet windows. Of these, 270 vary the rates of Video Conference (six, 3--70~Mb/s), Bursty UDP and IPERF UDP (five each, 5--70~Mb/s), and IPERF TCP (four, measured at 5--197~Mb/s). The other 78 cover Gaming, VoIP, and Video Streaming at fixed rates. For Dynamic Adaptive Streaming over HTTP (DASH), the configured rate describes the video profile; segments are downloaded in bursts.

\subsection{Wi-Fi Simulation Scenarios}\label{sec:wifi-scenarios}

The Wi-Fi simulations use a modified version of ns-3.40 with a privately maintained model of the IEEE 802.11 physical (PHY) and medium access control (MAC) layers. The Wi-Fi links use IEEE 802.11be Extremely High Throughput (EHT) at 5~GHz, a 160~MHz channel, modulation and coding scheme (MCS)~5, one spatial stream, and an 800~ns guard interval. All application flows use the Best Effort (BE) access category. Each simulation lasts 60~s, and applications start at 2~s.

Table~\ref{tab:scenarios} lists the downlink flows and rates using the application IDs in Table~\ref{tab:apps}. In S3--S6, BSS~1 carries the listed home-network flows, while a neighboring BSS~2 adds eight UDP uplink flows, each at 70~Mb/s, on the same channel. S7 covers all eight classes with nine downlink flows, each to a separate station. Its Video Conference, VoIP, and Gaming applications also generate uplink traffic (4, 0.10, and 0.11~Mb/s, respectively).

\begin{table*}[t]
\caption{Wi-Fi scenarios: downlink flow rates (Mb/s) by application ID from Table~\ref{tab:apps}.}
\label{tab:scenarios}
\centering\footnotesize
\setlength{\tabcolsep}{6pt}
\newcommand{\hd}[1]{\textbf{\begin{tabular}[b]{@{}c@{}}#1\end{tabular}}}
\begin{tabular}{@{}l l c c c c c c c c c r@{}}
\toprule
\hd{Scenario} & \hd{Network} & \hd{VP} & \hd{VC} & \hd{GM} & \hd{VS} & \hd{BT} & \hd{IT} & \hd{IU} & \hd{BU} & \hd{Flows} & \hd{Windows}\\
\midrule
S1 & One BSS & 0.10 & 5, 10 & 0.11, 0.11 & 20 & 15 & -- & 200, 200 & 100 & 10 & 2,699\\
S2 & One BSS & 0.10 & 5, 10 & 0.11 & 20 & 15 & unb. & 200, 200 & 100 & 10 & 2,691\\
S3 & Two BSSs\textsuperscript{*} & 0.10 & 5 & 0.11 & 20 & -- & -- & -- & -- & 4 & 1,080\\
S4 & Two BSSs\textsuperscript{*} & 0.10 & 5 & 0.11 & 20, 10 & -- & -- & -- & -- & 5 & 1,350\\
S5 & Two BSSs\textsuperscript{*} & 0.10 & 5 & 0.11 & 20 & -- & -- & -- & 300 & 5 & 1,346\\
S6 & Two BSSs\textsuperscript{*} & 0.10 & 5 & 0.11 & 20 & 15 & -- & -- & 300 & 6 & 1,552\\
\midrule
S7 & One BSS & 0.10 & 8 & 0.11 & 20, 10 & 50 & unb. & 20 & 10 & 9 & 2,430\\
\bottomrule
\end{tabular}
\par\vspace{2pt}
\parbox[t]{\textwidth}{\footnotesize Each value is the rate of one downlink flow; --: none; unb.: unbounded. \textsuperscript{*}BSS~1 is the home network; co-channel BSS~2 contributes eight UDP uplink flows at 70~Mb/s each. Flows: home-network downlink flows per run. Windows: evaluated windows across five simulation seeds. S1--S6 are development scenarios; S7 is the held-out test scenario.}
\end{table*}

Each Wi-Fi scenario uses five simulation seeds (five runs per scenario), giving 30 runs for S1--S6.

The AP-side extractor logs candidate statistics by flow and window. The first window covers 2--7~s; each flow is labeled by its destination station's application. Across 30 S1--S6 runs, 40 downlink flows yield 10,718 windows, 82 short of the 10,800 maximum (54 per flow per run), because congestion leaves some flows incomplete. S7 follows the same procedure.

\subsection{Implementation and Evaluation Protocol}\label{sec:evaluation-protocol}

For the Extreme Gradient Boosting (XGBoost) classifier \cite{xgboost}, hyperparameters are selected by grid search using the initial 29 features and only Ethernet training data. Four-fold cross-validation is performed using four Ethernet training captures, each generated with a different
simulation seed. In each fold, three captures are used for
training and one for validation, ensuring that overlapping
windows from the same capture are not shared between
training and validation. A fifth independent capture is
reserved exclusively for Ethernet testing. Hyperparameters are selected based on mean validation accuracy, while macro-F1 is also calculated. The selected settings are 400 boosting rounds, maximum depth six, learning rate 0.08, row and column subsampling of 0.85, $reg\_alpha=0.1$, and $reg\_lambda=1.0$. To isolate the effect of feature selection and construction, the same hyperparameters are used to train a separate classifier for each of the six feature sets (29, 28, 16, 12, 9, and 7 inputs), using all Ethernet training windows.

The seven-feature model also runs online with the Wi-Fi extractor. For comparison, all six classifiers are evaluated offline on the same logged Wi-Fi simulation windows, without further training.

The four evaluation sets are Ethernet training and test, Wi-Fi development (S1--S6), and Wi-Fi test (S7), with counts in Table~\ref{tab:apps}. S1--S6 guide feature design but not model training; S7 is excluded from both. Training accuracy measures fit to the training data.

For S1--S6, accuracy, class recall, and macro-F1 are calculated from all 10,718 windows pooled across 30 runs. Each window has equal weight, rather than each scenario. This summarizes the combined development scenarios. For the held-out S7 scenario, accuracy is reported as the mean $\pm$ sample standard deviation across five runs, each containing 486 windows. Because all five runs contain the same number of windows, mean accuracy equals accuracy over all 2,430 pooled S7 windows; the standard deviation describes variation across runs. S7 class recall and macro-F1 also use pooled windows. Consecutive five-second windows overlap by four seconds and are not independent observations.

\section{Results and Discussion}\label{sec:results}

\subsection{Effect of the Selection and Construction Stages}

\begin{figure}[!htbp]
\centering\includegraphics[width=\columnwidth]{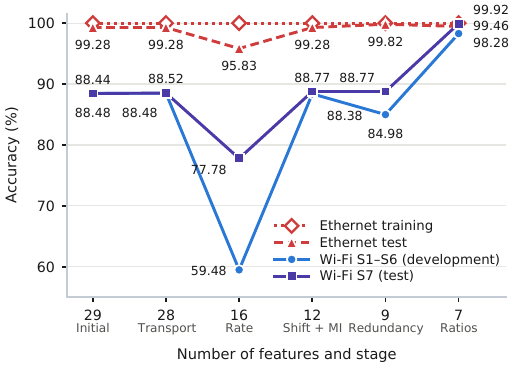}
\caption{Accuracy after each feature selection and construction stage (Section~\ref{sec:feature-procedure}; Ratios: feature construction) on the four evaluation sets of Section~\ref{sec:evaluation-protocol}. Ethernet training accuracy is 100\% at every stage. Where the two Wi-Fi lines overlap, the upper value is S7 and the lower value is S1--S6.}
\label{fig:accuracy}
\end{figure}

Fig.~\ref{fig:accuracy} shows the accuracy obtained with each successive feature set. We examine the main errors below.

Ethernet test accuracy is 99.28\% with 29 features and 99.46\% with 7 features, reaching its minimum after the rate-dependence filter.

Removing \feat{protocol} (29 to 28 features) leaves S1--S6 accuracy unchanged because the TCP flag count still distinguishes TCP and UDP. With 29 features, 932 Video Streaming windows are classified as IPERF TCP and 266 Bulk TCP windows as Video Streaming; results with 28 features are similar.

The rate-dependence filter (28 to 16 features) removes the TCP flag count and lowers accuracy on the Ethernet test set, S1--S6, and S7. In S1--S6, 3,092 Video Conference, IPERF UDP, and Bursty UDP windows are classified as IPERF TCP. These UDP classes use fixed training rates; IPERF TCP covers four. Higher-rate UDP traffic in S1--S6 and Video Conference traffic in S7 produce shorter mean inter-arrival times within the IPERF TCP training range.

The KS and MI criteria (16 to 12 features) remove \feat{mean_iat}, \feat{std_iat}, \feat{avg_min_pkt_size}, and \feat{directions}, restoring correct classification of these UDP flows.

Redundancy reduction (12 to nine features) misclassifies all 540 windows of the 300~Mb/s Bursty UDP flows in S5 and S6 as IPERF UDP. Their constant 1,474-byte packets at 40~\textmu s intervals resemble IPERF UDP rather than the variable sizes and gaps of Bursty UDP training traffic. With 12 features, TreeSHAP~\cite{lundberg2020} identifies \feat{mean_pkt_size} as the largest contribution toward Bursty UDP. It is removed as redundant with \feat{rms_pkt_size} in training, yet adding it back restores correct classification. Features that are redundant in Ethernet training may affect Wi-Fi classification differently.

Feature construction (nine to seven inputs) retains three direct statistics and introduces four ratio-based features. With this set, Bursty UDP is classified correctly and both Wi-Fi sets achieve their highest accuracy. The remaining errors are examined in Sections~\ref{sec:s1s6errors} and~\ref{sec:s7results}.

\subsection{Sensitivity to Selection Thresholds}

The same nine raw features are selected when each threshold is varied separately within the following ranges: rate correlation, 0.8326--0.9712 (chosen: 0.97); KS distance, 0.2690--0.3558 (0.30); MI ratio for $D>0.30$, 0.8964--1.1572 (1.00); minimum MI ratio, 0.5144--0.7768 (0.60); and redundancy correlation, 0.9416--0.9914 (0.98). The chosen rate threshold is close to the upper end of its range.

For joint threshold variation, we test one representative from each of $4\times9\times10\times5\times22=39{,}600$ regions; all retain the same nine raw features. Results are also unchanged across ten MI-estimator seeds. This establishes stability within the tested ranges, not global optimality.

\subsection{Classification Errors on S1--S6}\label{sec:s1s6errors}

With seven features, 165 IPERF TCP windows and ten Bulk TCP windows are classified as Video Streaming. This reduces IPERF TCP recall to 38.9\% (Table~\ref{tab:recall}).

\begin{table}[!htbp]
\caption{Recall (\%) per class, accuracy, Macro-F1 with 29 and 7  features.}
\label{tab:recall}
\centering\footnotesize
\setlength{\tabcolsep}{3pt}
\begin{tabularx}{\columnwidth}{@{}Y r r r r@{}}
\toprule
 & \multicolumn{2}{c}{\textbf{S1--S6}} & \multicolumn{2}{c}{\textbf{S7}}\\
\cmidrule(lr){2-3}\cmidrule(l){4-5}
\textbf{Class} & \textbf{29} & \textbf{7} & \textbf{29} & \textbf{7}\\
\midrule
Bulk TCP & 63.1 & 98.2 & 0.0 & 100\\
IPERF TCP & 93.0 & 38.9 & 99.3 & 99.3\\
Video Streaming & 49.8 & 99.7 & 98.3 & 100\\
Other five classes & 100 & 100 & 100 & 100\\
\midrule
Accuracy & 88.48 & 98.28 & $88.44\pm0.27$ & $99.92\pm0.18$\\
Macro-F1 & 0.840 & 0.937 & 0.831 & 0.999\\
\bottomrule
\end{tabularx}
\par\vspace{2pt}
\parbox[t]{\columnwidth}{\footnotesize Other five classes: VoIP, Video Conference, Gaming, IPERF UDP, and Bursty UDP. S1--S6: pooled metrics over 30 runs. S7: accuracy mean $\pm$ sample standard deviation across five equal-sized runs; its mean equals pooled accuracy. Class recall and macro-F1 use pooled windows.}
\end{table}

For models using 29 to 12 features, TreeSHAP identifies observed duration as one of the two inputs most strongly favoring IPERF TCP. In Ethernet training, IPERF TCP windows span five seconds, while Video Streaming windows are shorter. On Wi-Fi, most windows of both classes span nearly five seconds. With these feature sets, about half of the Video Streaming windows are classified as IPERF TCP.

With seven features, nearly all Video Streaming windows are classified correctly, but many IPERF TCP windows are classified as Video Streaming. Because S1--S6 contain about seven times more Video Streaming windows than IPERF TCP windows, overall accuracy increases despite the lower IPERF TCP recall.

In S2, IPERF TCP and Video Streaming have similar packet sizes, making them harder to distinguish. In Ethernet training, IPERF TCP sends continuously (\feat{activation_fraction} $\geq0.9996$), while Video Streaming alternates downloads and idle periods ($\leq0.68$). High load introduces longer IPERF TCP gaps while making Video Streaming more continuous. Their timing becomes less distinct, and almost all IPERF TCP windows with \feat{activation_fraction} below 0.98 are classified as Video Streaming.

\subsection{Results on Test Scenario S7}\label{sec:s7results}

IPERF TCP is unbounded in S2 and S7, but S7's lower competing load allows higher throughput and shorter AP delay (Table~\ref{tab:s2s7}). Only two S7 windows are misclassified, both as Video Streaming. They are the first two IPERF TCP windows of one run and include a startup pause that lowers \feat{activation_fraction} below 0.98.

\begin{table}[!htbp]
\caption{Traffic conditions in S2 and S7, and IPERF TCP recall with seven features.}
\label{tab:s2s7}
\centering\footnotesize
\setlength{\tabcolsep}{3pt}
\begin{tabularx}{\columnwidth}{@{}Y r r@{}}
\toprule
 & \textbf{S2} & \textbf{S7}\\
\midrule
Downlink load requested by the other flows (Mb/s) & $\approx$550 & $\approx$120\\
IPERF TCP throughput (Mb/s) & 12 & 197\\
Mean downlink delay of IPERF TCP at the AP (ms) & 147 & 8\\
IPERF TCP windows with \feat{activation_fraction} $<0.98$ & 63\% & 0.7\%\\
Median \feat{activation_fraction} of Video Streaming & 0.97 & 0.72\\
IPERF TCP recall (\%) & 38.9 & 99.3\\
\bottomrule
\end{tabularx}
\par\vspace{2pt}
\parbox[t]{\columnwidth}{\footnotesize In S2, the AP delivers about 370~Mb/s in total. In Ethernet training, the median \feat{activation_fraction} of Video Streaming is 0.64.}
\end{table}

With 29 features, all 270 Bulk TCP windows are classified as IPERF TCP. The higher Bulk TCP rate places its rate-dependent features within the IPERF TCP training range. Removing them is insufficient: most errors persist with 16, 12, and nine features. The longest gap in a Bulk TCP window decreases from 1.2~ms in training to 0.24~ms in S7, close to the IPERF TCP training value of 0.22~ms. TreeSHAP identifies \feat{duration}, \feat{max_iat}, and \feat{std_pkt_size} among the inputs favoring IPERF TCP.

Seven features classify all Bulk TCP windows correctly. Its maximum packet size remains 1,475~bytes, compared with at most 592~bytes for IPERF TCP and Video Streaming. Contributions from \feat{max_pkt_size}, \feat{min_max_ratio}, and \feat{switching_ratio} outweigh \feat{activation_fraction}, which still favors IPERF TCP. The nine-feature model already includes the same size statistics but still misclassifies most Bulk TCP windows. Among the evaluated models, all these windows are classified correctly only after feature construction.

Adaptive video traffic can alternate downloads and idle periods, use separate audio and video chunks, and vary segment size and bitrate~\cite{rao2011,akhshabi2011,loh2021,juluri2016,biernacki2025}. These choices affect packet timing with load and playback state. Our evaluation uses one simulated DASH client and specific traffic generators, limiting generalization to other implementations and conditions.

\section{Conclusion}\label{sec:conclusion}




Our study shows that traffic classifiers trained only on Ethernet data can achieve high classification accuracy on Wi-Fi through careful feature selection and construction. We propose a procedure to reduce the large initial feature set to a compact set of seven flow features that capture packet-size and timing patterns while avoiding features that depend too strongly on traffic rate or differ sharply between Ethernet and Wi-Fi. In the ns-3 simulations, this compact feature space improved accuracy on the held-out Wi-Fi scenario (S7) from about 88\% to about 99.9\%, while Ethernet accuracy remained high at about 99.5\%. Our conclusion is that careful feature selection and construction can help Ethernet-trained models transfer to Wi-Fi without using Wi-Fi data for classifier training. 
Future work will evaluate whether the selected features remain suitable under different traffic loads, streaming implementations, and real network conditions.

\section*{Acknowledgment}

This work was supported by the EU Horizon Europe Program in the frame of Marie Skłodowska-Curie Actions Doctoral Networks (MSCA-DN) project OWIN6G (Optical and wireless sensors networks for 6G scenarios) under Agreement No. 101119624.

\bibliographystyle{IEEEtran}
\bibliography{references}

\end{document}